\documentclass[12pt,preprint]{aastex}
\newcommand{\HI}{H~{\sc i}} 

\newcommand{\kms}{${\rm km~s^{-1}}$}

\slugcomment{Accepted for publication in ApJ Letters}
\shortauthors{McCLURE-GRIFFITHS ET AL.} 

\begin{document} 

\title{Atomic Hydrogen in a Galactic Center Outflow}

\author{N.\ M.\ McClure-Griffiths,\altaffilmark{1} 
J.\ A.\ Green,\altaffilmark{1}  
A.\ S.\ Hill,\altaffilmark{1}  
F.\ J.\ Lockman,\altaffilmark{2} 
J.\ M.\ Dickey,\altaffilmark{3} 
B.\ M.\ Gaensler,\altaffilmark{4} \&
A.\ J.\ Green\altaffilmark{4} }
\altaffiltext{1}{Australia Telescope National Facility, CSIRO
  Astronomy \& Space Science,  Marsfield NSW 2122, Australia;
  naomi.mcclure-griffiths@csiro.au}
\altaffiltext{2}{National Radio Astronomy Observatory, Green Bank WV
  24944, USA}
\altaffiltext{3}{School of Physics and Mathematics, University of
  Tasmania, TAS 7001, Australia}
\altaffiltext{4}{Sydney Institute for Astronomy, School of Physics,
  The University of Sydney, NSW 2006, Australia}

\begin{abstract}
  We describe a population of small, high velocity, atomic hydrogen
  clouds, loops, and filaments found above and below the disk near the
  Galactic Center.  The objects have a mean radius of 15 pc, velocity
  widths of $\sim 14$ \kms\ and are observed at $|z|$ heights up to
  700 pc.  The velocity distribution of the clouds shows no signature
  of Galactic rotation.  We propose a scenario where the clouds are
  associated with an outflow from a central star-forming region at the
  Galactic Center.  We discuss the clouds as entrained material
  traveling at $\sim 200$ \kms\ in a Galactic wind.
\end{abstract}


\section{Introduction}
\label{sec:intro}
Many galaxies with active star formation in their central regions have
 large-scale Galactic winds.  These are thought to be
responsible for circulating enriched gas to the halo of their host galaxy and even to the
intergalactic medium.  The galaxy 
M82 has not only hot, ionized gas in its outflow, but also cool
H$\alpha$ emitting filaments \citep{bland88,shopbell98}, and cold
molecular gas \citep[e.g.][]{nakai87,taylor01}.  Simulations
of starburst winds have shown that dense, molecular gas may be
entrained in outflowing winds  producing cold clouds and filaments
at high latitudes around the center of a galaxy \citep[e.g.][]{cooper08,melioli13}.

There is increasing evidence that the Milky Way may also have a
large-scale nuclear wind.  \citet{bland-hawthorn03} linked the
Galactic Center lobe and North Polar Spur to a bipolar wind traced by
infrared emission in the inner $\sim 2\arcdeg$ of the Galaxy.  In
gamma-ray emission there are two very large lobes extending to
$|b|\sim60\arcdeg$, centered on the Galactic Center \citep{su10}.
Recently \citet{carretti13} identified counterparts to the Fermi
bubbles in radio polarized emission and suggested that the lobes are
formed through the effects of massive star formation, which drives a
Galactic wind at $\sim 1100$ \kms.  There are few direct tracers of
the velocity of the putative nuclear wind of the Milky Way, and little
evidence of associated cold gas.  In fact, the inner 3 kpc of the
Milky Way shows a distinct lack of diffuse \HI\ at $|z|<400$ pc, which
\citet{lockman84} attributed to a Galactic wind removing \HI\ except
in the narrow disk \citep[e.g.][]{bregman80}.  The most
convincing kinematic information comes from \citet{keeney06} who found
high velocity UV absorption along two sight-lines indicating a maximum
outflow velocity of 250 \kms.

We have discovered a population of compact, isolated clouds in
the Australia Telescope Compact Array (ATCA) \HI\ Galactic Center
Survey that may trace neutral gas in a wind.  

\section{A population of Clouds}
\label{sec:clouds}

The ATCA \HI\ Galactic Center Survey \citep{mcgriff12} covers Galactic
longitudes $-5\arcdeg \leq l \leq +5\arcdeg$, and Galactic latitude
$-5\arcdeg \leq b \leq +5\arcdeg$ over the velocity range $-309~{\rm
  km~s^{-1}} \leq V_{LSR} \leq +349$ \kms\ with a velocity resolution
of 1 \kms,  an angular resolution of $145\arcsec$ and a mean rms \HI\
brightness temperature of $\sigma_{T_b}=0.7$ K.

\begin{figure}
\centering
\plotone{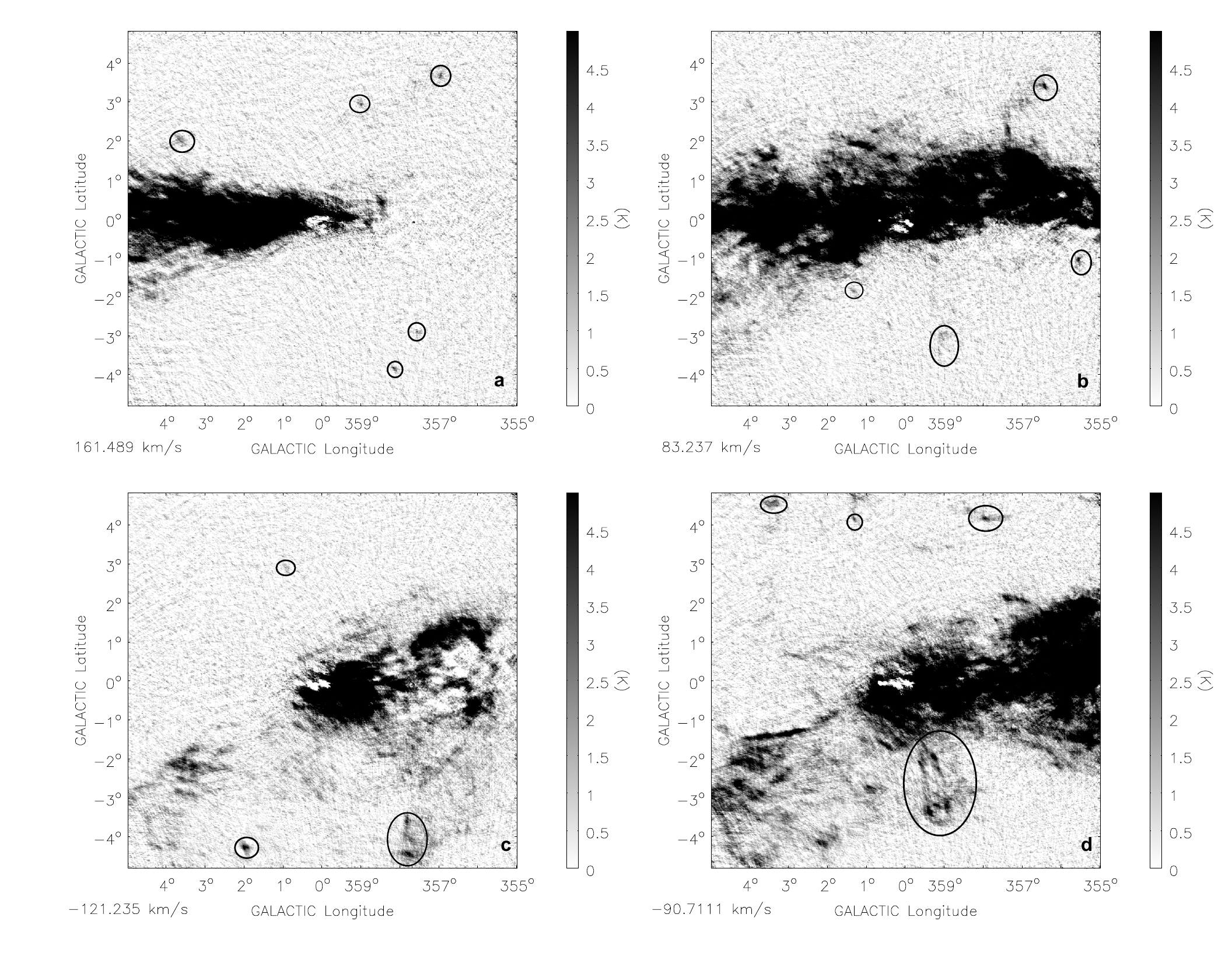}
\caption[]{\HI\ images at (a) $v=161.5$ \kms,  (b) $v=83.2$ \kms, (c)
  $v=-121.2$ \kms, and $v=-90.7$ \kms.  Catalogued clouds whose central velocity
  lies within $5$ \kms\ of the shown velocity are circled.  
\label{fig:example}}
\end{figure}

In the ATCA survey data for the central $1500~{\rm pc} \times
1500~{\rm pc}$ of the Galaxy there is a population of compact \HI\
clouds that are isolated both spatially and in velocity from the bulk
of Galactic \HI\ emission.  Clouds were selected to be relatively
compact ($\lesssim 0.5\arcdeg$), discrete from other emission, and
with brightness temperatures greater than $2\sigma_{T_b}$ over more than
five spectral channels. Typically the clouds are only just spatially
resolved.  However, there are some more extended filamentary features,
which we also refer to generally as clouds.  Example
clouds are shown in Figure \ref{fig:example}.  We have found a total
of 86 clouds between $-209~{\rm km~s^{-1}} \leq V_{LSR} \leq 200~{\rm
  km~s^{-1}}$.

\begin{figure}
\centering
\plotone{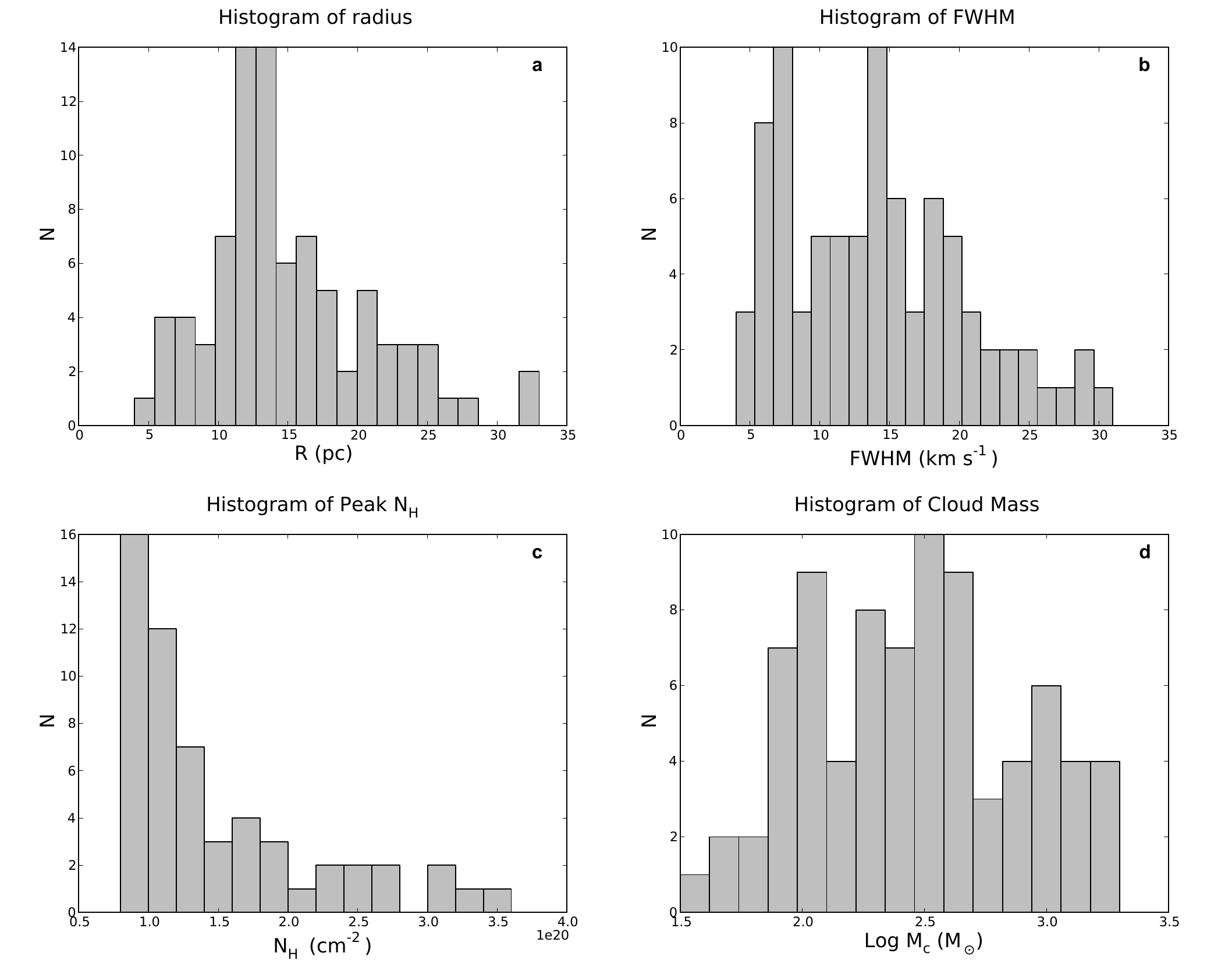}
\caption[]{Histograms of: (a) the mean radius of individual clouds, (b) the
  velocity FWHM of the clouds,  (c) the peak column density, $N_H$,
  of the clouds and  (d) the total mass of the clouds.
\label{fig:histos}}
\end{figure}

For each cloud we measure the central velocity, $V_{LSR}$, fitted peak
brightness temperature after baseline removal, $T_b$, velocity
full-width half-max $\Delta v$, mean radius, $r$, assuming a distance
of $R_{\sun} = 8.5$ kpc, peak column density, $N_{H\,max}$, and cloud
mass, $M_c$.  The cloud properties are given in
Table~\ref{tab:catalogue}.  The mean radius, $r = \sqrt{A/\pi}$, is
calculated from the area of the cloud, $A$, where $A$ is the number of
pixels in the integrated intensity image with $N_H > 0.4N_{H\,max}$.
This is clearly a very rough estimate as many clouds are not well
represented by an ellipse.  The \HI\ masses are calculated as the sum
of all pixels greater than $40$\% $N_{H\,max}$ assuming a distance of
8.5 kpc.  We quote \HI\ masses assuming the gas is optically thin.
The full width half max of the \HI\ line (FWHM) is shown in
Figure~\ref{fig:histos}, along with histograms of mass, radius and
column density for these clouds.  The median, minimum and maximum
values for the population are given in Table \ref{tab:props}.  There
are no strong correlations of cloud radius, FWHM or mass with $z$
height.  The median line width corresponds to $T_k \equiv 22\Delta v^2
\leq 4000$ K.  Assuming that the unresolved clouds are roughly
spherical we estimate an average \HI\ number density,
$\left<n_c\right> \sim N_H / (2r) = 1~{\rm cm^{-3}}$.

There are several larger features, such as the loop at $(l,b)=
(359\arcdeg,-3\arcdeg$) (Figure \ref{fig:example}d), which appear to
cohesive over many degrees. These features have also larger velocity widths
and velocity gradients.  The $(l,b)= (359\arcdeg,-3\arcdeg$) loop is
particularly interesting because it appears to point directly back to
the Galactic Center.

\subsection{Cloud Kinematics}
\label{subsec:distribution}
\begin{figure}
\centering
\includegraphics[width=5in]{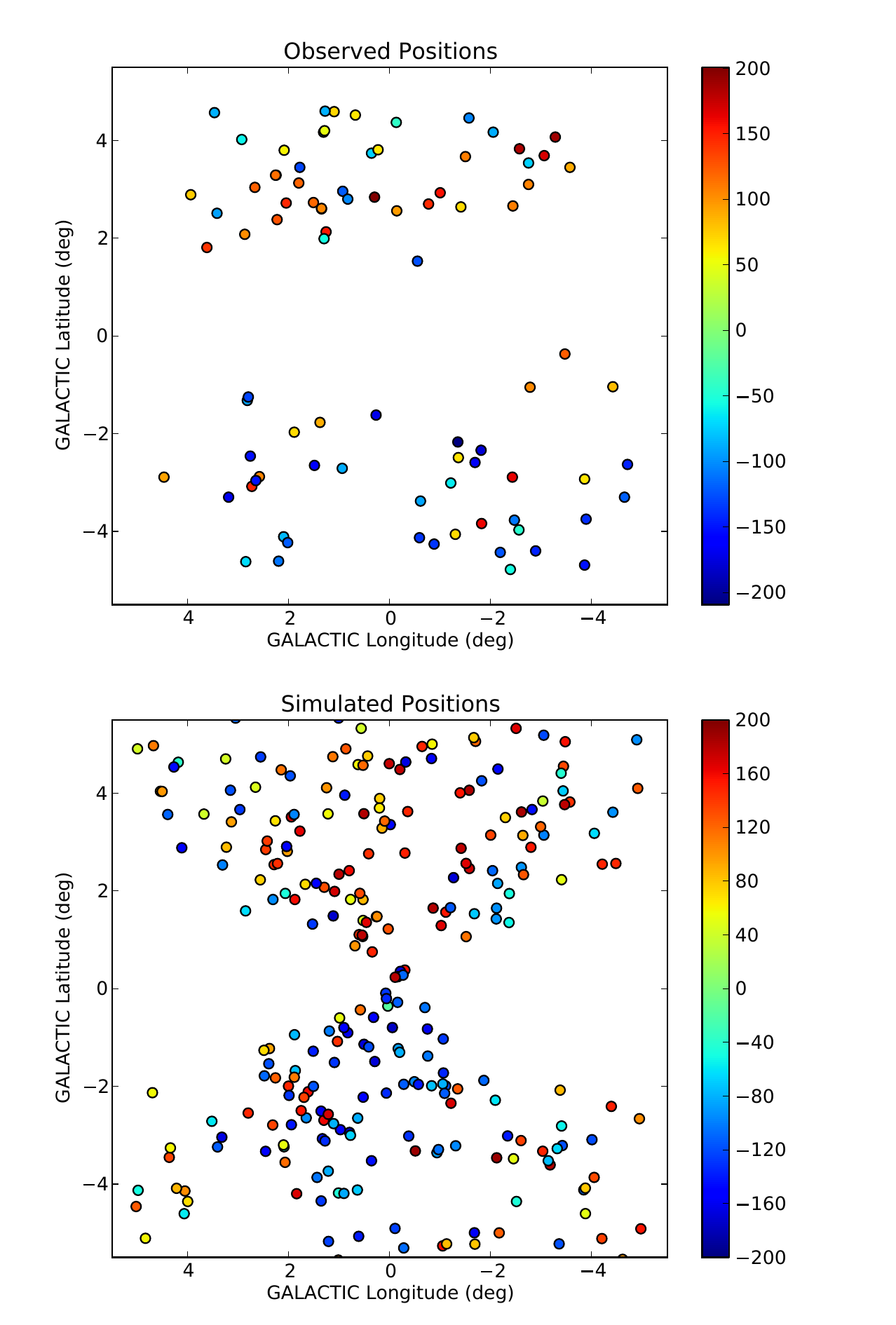}
\caption[]{Distribution of clouds around the Galactic
  Center found in ({\bf a}):  the ATCA Galactic Center survey and  ({\bf b}): a simulated wind of
  velocity 190 \kms.  The color of the symbols is
  determined by the central velocity of the sources.  The LSR velocity
  of the simulated clouds is calculated from Equation \ref{eq:vw}.  
  \label{fig:dist}}
\end{figure}

The location of the clouds, color-coded by their LSR
velocities, is shown in Figure~\ref{fig:dist}a.  
The clouds clearly do not follow Galactic rotation.  The velocity
projection of an object at $l,b$ onto the local standard of rest can be written 
\begin{equation}
V_{LSR} = \left[R_0 \sin{l} \left(\frac{V_{\theta}}{R} - \frac{V_{\sun}}{R_{\sun}}\right) - V_R
\cos{(l+\theta)}\right]\, \cos{b} + V_z \sin{b},
\end{equation}
where $\theta$ is defined around the Galactic Center in a clockwise
direction from the Sun-Galactic Center line.  The velocity terms
$V_R$, $V_{\theta}$ and $V_z$ are the radial, azimuthal and vertical
velocities.  In circular Galactic rotation $V_R$ and $V_z$ are zero
and the $\sin{l}$ term produces a change in sign of $V_{LSR}$ across
$l=0\arcdeg$.  The absence of this signature in the cloud distribution
suggests that the kinematics of the clouds are not dominated by
circular Galactic rotation.  Because the clouds are at high velocities
and do not show the signature of Galactic rotation we assume that the
clouds are located at the Galactic Center.  
 
Clouds above the Galactic plane have predominantly positive velocities
while clouds below the Galactic plane have predominantly negative
velocities.  Above the plane there are 16 clouds at negative
velocities and 29 clouds at positive velocities.  Below the plane
there are 28 clouds at negative velocities and 13 clouds at positive
velocities.  The mean LSR velocity at positive latitudes is $40.7$
\kms\ and the mean velocity at negative latitudes is $-52.8$ \kms.
This is unlikely to be the effect of the $V_z$ term in Equation 1, as $\sin{b}$ is very
small and would require $V_z$ values of $5000 - 10000$ \kms\ to
explain the observed range of LSR velocities.

This kinematic asymmetry   is likely a product of 
confusion with unrelated  \HI\ emission, which effects our
ability to find isolated clouds.  It is well known that most of the
\HI\ emission around the Galactic center at $v_{LSR}>100$ \kms\ is at
$l > 0\arcdeg$ and $b < 0\arcdeg$ and similarly the emission at
$v_{LSR}<-100$ \kms\ is at $l < 0\arcdeg$ and $b > 0\arcdeg$ (Burton
\& Liszt 1978 \nocite{burton78}, see also
Figure 4 of McClure-Griffiths et al.\ 2012).  Consequently, we are
not likely to find many high velocity clouds isolated in $l$-$b$-$v$ space with
positive velocities at $b < 0\arcdeg$ or with 
negative velocities at $b > 0\arcdeg$.

\subsection{Simulated population}
\label{subsec:simpop}
To understand the selection effects on the distribution of observed
cloud velocities we have simulated clouds in an outflowing wind.  We
have populated the interior of two cones having an opening angle,
$\alpha$, centered on the Galactic Center and extending to a height of
1.5 kpc with 1000 randomly distributed clouds.  Each cloud was
assigned an LSR velocity based on its location, a constant conical
wind of velocity, $V_w$, and the expression similar to that derived by
\citet{keeney06}\footnote{This is a near reproduction of Equation 1
  from \citet{keeney06} except the second term is raised to the power
  of $1/2$, correcting a typographical error in Keeney et al.}
\begin{equation}
V_{LSR} = V_w \,\left[\frac{\rho}{r} \left(1 - \cos^2{b}
  \cos^2{l}\right) \mp \cos{b}\cos{l} \left[ 1 -
  \frac{\rho^2}{r^2}\left(1 - \cos^2{b}\cos^2{l}\right)\right]^{1/2}
\right] -
220 \sin{l} \cos{b}. 
\label{eq:vw}
\end{equation}
Here $l$ and $b$ are the Galactic coordinates,  $\rho$ is
the distance along the line of sight to a cloud, $r$ is  the
cloud's distance from the Galactic Center,  and  $r^2 =
R_{\sun}^2 - 2 \rho R_{\sun}\cos{b}\cos{l} + \rho^2$.  The second term is negative for  
objects on the near side of the Galactic Center and 
positive for objects on the far side of the Galactic
Center from the Sun.  We assumed the IAU standard  $R_{\sun}=8.5$ kpc.

Each simulated cloud was compared to the \HI\ actually detected in the
ATCA survey at its $l$ and $b$.  If the cloud lay at a velocity where
there was no significant \HI\ emission ($<4\sigma_{T_b}$) it was
assumed to have been "detected".  Models covered wind velocities,
$V_w$, between 100 \kms\ and 800 \kms.  Wind velocities $\lesssim 150$
\kms\ are unable to produce high velocity clouds as observed, whereas
large wind velocities produce a cloud distribution whose mean velocity
is higher than observed.  Using a Kolmogorov-Smirnov (K-S) test, and
separating the data into positive and negative latitude regions, we
estimated the probability that the velocities of the observed clouds
could be drawn from the same distribution as the simulated clouds.
The K-S test $p$-values as a function of $V_w$ showed a peak
near $V_w\sim 190$ \kms\ for positive latitudes and near $V_w \sim
220$ \kms\ for negative latitudes, but at a lower confidence level.
For all $V_w$ the simulated velocities were more consistent with the
observed velocity distribution at positive latitudes than at negative
latitudes.  Adopting a threshold of $p < 0.05$, the K-S test implies
that wind velocities greater than 270 \kms\ and less than 150 \kms\
are not consistent with the observed cloud velocities.

The observed cloud longitudes and latitudes set a lower limit on the
cone opening angle, $\alpha$.  Values $\alpha \lesssim 3\pi/4$ radians
are too narrow to produce intermediate latitude ($|b|\approx
2\arcdeg$) clouds at $|l| > 4\arcdeg$. The upper limit to $\alpha$ not
well constrained as there is coupling between the angle and the
assumed wind velocity.  We experimented with several angles, including
a fully spherical distribution, and found the results were best
matched to $\alpha=3\pi/4$ radians.

The resulting ``detected'' clouds for a wind velocity of $V_w=190$
\kms\ and $\alpha = 3\pi/4$ are shown in Figure
\ref{fig:dist}b.  Although the mean velocity of the initial distribution at both
positive and negative latitudes was close to zero, because of the
selection effects from blending with unrelated \HI\ emission, the
velocity distribution of the detected clouds shows the same preference
for positive velocities at $b>0\arcdeg$ and negative velocities at
$b<0\arcdeg$ observed in the real clouds.  From 50 realizations of the
simulation with $V_w = 190$ \kms\ we find mean and median positive
latitude velocities of $52$ \kms\ and $88$ \kms, and mean and median
negative latitude velocities of $-4$ \kms, and $-60$ \kms.  The real
data have mean and median values, respectively, of $40.7$ \kms\ and
$75.4$ \kms\ at $b>0\arcdeg$, and $-52.8$ \kms\ and $-93.3$ \kms\ at
$b<0\arcdeg$.  The velocity discrepancies at negative latitudes may
indicate a higher wind velocity below the Galactic plane or a wind
that does not point directly down.

The simulation produces a large number of clouds at small longitudes
($-1\arcdeg \lesssim l \lesssim +1\arcdeg$) and small velocities
$-70~{\rm km~s^{-1}} \lesssim V_{LSR} \lesssim +70~{\rm km~s^{-1}}$ at
all latitudes.  These clouds are collimated along the wind axis
perpendicular to the Galactic plane.  In most cases we expect these to be blended with
unrelated low velocity \HI\ emission.  However, examining the
observed \HI\ data
cube we do find significant structure collimated perpendicular to the
Galactic plane at $-1\arcdeg \lesssim l \lesssim +1\arcdeg$ with
velocities $|V_{LSR}| \sim 50 - 70~{\rm km~s^{-1}}$.  These features
do not meet our criteria of being discrete from other emission and
were therefore not catalogued, but may reasonably be part of the same
wind structure.

The simulation thus approximately reproduces some of the key features
of the observed distribution.  The observations are consistent with a
outflowing conical wind of $\sim 200$ \kms.  We are also able to say
with confidence that the observed velocity distribution of clouds with
positive velocities at $b>0\arcdeg$ and negative velocities at
$b<0\arcdeg$ is caused by confusion with unrelated \HI.

\section{Discussion and Conclusions}
\label{sec:disc}
The recent discovery of the Fermi bubbles \citep{su10} in $\gamma$-ray
emission and their apparent counterparts in polarized radio continuum
\citep{carretti13} has fueled speculation that there is either a
powerful jet from the Galactic Center or there has been a past burst
of star formation.  In explaining the polarized radio continuum
structure of the Fermi bubbles, \citet{carretti13} argue that the
bubbles are consistent with a past burst of star formation and confirm
the \citet{bland-hawthorn03} estimate that the bubbles required an
injection of $\sim 10^{55}~{\rm erg}$.  They estimate that the
interiors of the Fermi bubbles are filled with hot plasma of
temperature $T_w\sim 10^7$ K with a density, $n_w\sim 9\times
10^{-3}~{\rm cm^{-3}}$, traveling at a velocity of 1100 \kms.

The median radius, column density and velocity width of the Galactic
Center \HI\ are very similar to the lower halo tangent point clouds
identified by \citet{ford10} and \citet{lockman02a}.  \citet{ford10}
suggested that halo clouds are the remnants of supershells and
superbubbles pushed beyond the Galactic disk through the stellar winds
and supernovae of massive stars, but remaining dynamically linked to
the Galactic disk.  Our Galactic Center clouds have higher velocities
than the \citet{ford10} clouds.  The formation mechanism for the
Galactic Center clouds, which determines their dynamics, is therefore
probably more energetic than for the halo clouds.  This is not
surprising, the star formation rate in the inner 100 pc of the Galaxy
is $0.1~{\rm M_{\sun}~yr^{-1}}$ \citep[reviewed by][]{crocker12},
whereas the star formation rate at the end of the Galactic bar where
many halo clouds were found ($l \approx30\arcdeg$) is $\sim
10^{-4}~{\rm M_{\sun}~yr^{-1}}$ \citep{veneziani13}.

In spite of the extreme star formation rate in the very active region
of the Galactic Center and estimated high temperatures in the wind,
cold clouds are not unexpected.  Simulations by \citet{cooper08} of
starburst driven Galactic winds show compact clouds, filaments, and
complex structures of
density, $n\sim 10~{\rm cm^{-3}}$ and temperature, $T\sim 10^4~{\rm
  K}$ entrained in the inner 400 pc of a starburst wind.  Similarly,
the \citet{melioli13} simulations of the M82 wind produced a wealth of
features at $T\sim 10^4~{\rm K}$ with sizes $20-300$ pc and densities in
the range $10^{-1}-10~{\rm cm^{-3}}$.  The velocity of this cold
phase of the outflow is typically 10 - 30\% of the velocity of the hot
phase, or 100 - 800 \kms, for the
\citet{cooper08} and \citet{melioli13} models. \citet{melioli13} also found
that in order for the wind to contain a rich filamentary structure of
cool gas the starburst needed to occur over a relatively short
timescale of a few million years.

Our wind velocity modeling produces a velocity of the cold component
of the wind that is similar to that observed in the H$\alpha$ emitting
filaments of M82 \citep{shopbell98} and predicted by hydrodynamic
simulations \citep{cooper08,melioli13}.  We can estimate the velocity of
the hot component using the relationship for the ratio of the terminal
velocity of the cold clouds, $V_{t}$, to the hot wind, $V_{hw}$, for a
cloud of density, $n_c$, in a wind of density, $n_w$,
\citep{martin05}
\begin{equation}
\frac{V_t}{ V_{hw}} \simeq \left[ \frac{3 \, n_w}{2\, n_c}\right]^{1/2}.
\end{equation}  
Assuming $V_t = V_w=190$ \kms, as derived in \S\ref{subsec:simpop}, we
estimate the velocity of the hot wind as $V_{hw} \sim 1600$ \kms,
which is consistent with the hot wind velocity of 1100 \kms\ estimated by \citet{carretti13}.

Presuming that the clouds are launched from very near the galactic
plane, to reach $4\arcdeg$ latitude in a $\sim 200$ \kms\ wind
takes at least 3 Myr.  If the clouds are entrained in a hot wind then
they will be subject to a number of destructive effects.  First, they
will evaporate.  The evaporation rate, $\dot{m}$, for an unmagnetized
15 pc cloud in a $10^7$ K plasma with a density of $n_w\sim 9\times
10^{-3}~{\rm cm^{-3}}$ is $\dot{m}\sim 1\times 10^{-3}~{\rm
  M_{\sun}~yr^{-1}}$, assuming saturated evaporation \citep{cowie81b}.
This implies a typical survival time of a few $ \times 10^5~{\rm yr}$,
which is less than the time for a cloud to reach a $z$ height of 400
pc in a $200$ \kms\ wind.  However, the evaporation timescale is
highly sensitive ($\propto T^{5/2}$) to the assumed temperature of the
plasma surrounding the clouds.  Hydrodynamic models of winds show that
the temperature may be anywhere in the range ${\rm few} \times
10^{6-7}$ K, it therefore seems possible that the clouds could survive
for up to $\sim 1$ Myr.  Also important to the clouds' lifetimes are
the cloud crushing time, due to the wind shock, and the cloud cooling
time.  The relative timescales for these two effects gives an estimate
of whether the cold, dense clouds would be protected by an enveloping
radiative shock \citep{cooper08}.  The cooling time for a cloud of
density $n_c$ of shocked temperature $T_{cl,sh} = 3/16 \, T_{w}
(n_{w}/n_c)$ is on the order of a few hundred years.  This is much
less than the cloud crushing time, $t_{cc} \approx (n_c/n_w) (r_c/v_w)
\sim 4~{\rm Myr}$ \citep{cooper08}, which suggests that a cloud could
survive for some time in the hot wind.

The total mass of \HI\ in the observed clouds is only $\sim 4 \times
10^4~{\rm M_{\sun}}$.  The mass loss rate, assuming the oldest cloud
is $\sim 4$ Myr, is $0.01~{\rm M_{\sun}~yr^{-1}}$.  At this rate it is
unlikely that this outflow will have a significant impact on the $\sim
3\times 10^7~{\rm M_{\sun}}$ reservoir of cold gas estimated for the
100 pc ring in the Central Molecular Zone \citep{molinari11}.  We
estimate the work required to accelerate these clouds to their current
velocities; for a $270~{\rm M_{\sun}}$ cloud traveling at $200~{\rm
  km~s^{-1}}$, the kinetic energy is $\sim 1\times10^{50}~{\rm ergs}$,
and the kinetic energy of the most massive cloud in our sample is
$\sim 8\times 10^{50}~{\rm ergs}$. The total estimated kinetic energy
for all clouds traveling at a velocity of $200$ \kms\ is $\sim 1.5
\times 10^{52}$ ergs.  Assuming the ratio of ratio of kinetic energy
of the \HI\ clouds to the total input wind energy is $\sim 0.2$
\citep{weaver77}, as often assumed for superbubbles, the required wind
energy input was $\sim 8 \times 10^{52}$ ergs over the past $10^6$
yrs.  The wind luminosity is therefore $\sim 2 \times 10^{39}~{\rm
  erg~s^{-1}}$, which is easily supplied through supernovae and
stellar winds in the GC region with a star formation rate of $\sim 0.1
~{\rm M_{\sun}~yr^{-1}}$ \citep{crocker12}.

\section{Summary}
We have identified a population of compact, cool, $T<4000$ K, \HI\
clouds around the Galactic Center.  Although physically similar in
terms of size, temperature and column density to other cold \HI\
clouds in the lower halo of the Galaxy, the Galactic Center \HI\
clouds have unique kinematics and do not follow Galactic rotation.  We
suggest that these clouds are associated with a wind from the Galactic
Center and are the remnants of cold, dense gas entrained in the
outflowing wind.  Using a simple model for a Galactic wind with a
velocity of $\sim 200$ \kms\ we are able to roughly reproduce the
observed velocity distribution of the clouds.  This value, and
hydrodynamic simulations of a starburst wind, which find that the
velocity of the cold phase gas is 10 - 30\% of the velocity of the hot
phase \citep{cooper08,melioli13}, the hot wind emanating from the
Milky Way's Center must have a velocity of $\sim 600 - 1900$ \kms.  We
estimate the luminosity required to drive the clouds is $\sim 2 \times
10^{39}~{\rm erg~s^{-1}}$, which is easily supplied by stellar winds
and supernovae in the Galactic Center region over the past $\sim 2
\times 10^{6}$ years.

\acknowledgements We are grateful to R.\ Crocker for useful comments
on an early draft of this manuscript. The ATCA is part of the
Australia Telescope which is funded by the Commonwealth of Australia
for operation as a National Facility managed by CSIRO.
\bibliographystyle{apj} 


\begin{deluxetable}{lcccllcll}
\tabletypesize{\footnotesize}
\tablecaption{Compact clouds catalogue
\label{tab:catalogue}}
\tablehead{
\colhead{Name} & \colhead{$l$} & \colhead{$b$} & \colhead{$V_{LSR}$} & \colhead{Peak $T_b$} & \colhead{$\Delta v$}  &  \colhead{$N_{H\,max}$}  &  \colhead{$M_c$} & \colhead{$r$} \\ 
  \colhead{}& \colhead{(deg)} & \colhead{(deg)} & \colhead{(${\rm km~s^{-1}}$)} & \colhead{(K)} & \colhead{(${\rm km~s^{-1}}$)}  & \colhead{($\times 10^{19}~{\rm cm^{-2}}$)}  &  \colhead{(M$_{\sun}$)} & \colhead{(pc)} }
\startdata
 G0.2+3.8+65 	 & $0.23$ &  $3.81$  &  $63.7$ & $4.4$  & $5.9$  & $5.7$  & $25$ & $6.6$\\ 
 G0.2-1.6-170 	 & $0.27$ &  $-1.62$  &  $-169.6$ & $4.9$  & $17.5$  & $17.6$  & $2190$ & $32.9$\\ 
 G0.3+2.8+195 	 & $0.30$ &  $2.84$  &  $200.8$ & $1.4$  & $28.6$  & $9.7$  & $220$ & $12.2$\\ 
 G0.3+3.7-82 	 & $0.36$ &  $3.74$  &  $-74.3$ & $3.7$  & $13.7$  & $11.1$  & $706$ & $22.4$\\ 
 G0.7+4.5+61 	 & $0.68$ &  $4.52$  &  $63.3$ & $9.6$  & $5.3$  & $10.8$  & $287$ & $14.2$\\ 
\enddata
\tablecomments{Table \ref{tab:catalogue} is published in its entirety
  in the electronic edition of the {\it Astrophysical Journal
    Letters}.  A portion is shown here for guidance regarding its form
  and content.}
\end{deluxetable}

\begin{deluxetable}{llll}
\tablecaption{Cloud population properties
\label{tab:props}}
\tablehead{
\colhead{Property} & \colhead{Median} & \colhead{Min} & \colhead{Max} }
\startdata
$r~{\rm (pc)}$ & $15.0$ & $3.5$ & $32.9$ \\
$\Delta v~{\rm (km~s^{-1})}$ & $13.6$ & $3.0$ & $30.9$\\
$T_b~{\rm (K)}$ & $3.8$ & $1.4$ & $14.2$ \\
$N_{H\,max}~{\rm (10^{19}~cm^{-2})}$ &  $9.9$&$1.0$& $34.9$\\
$M_c~{\rm (M_{\sun})}$ & $270$ &$5$ & $2.1\times 10^3$ \\ 
\enddata
\end{deluxetable}

\end{document}